\documentclass[prl,nofootinbib]{revtex4}
\usepackage{graphicx}
\usepackage{latexsym}
\def\be{\begin{equation}}
\def\ee{\end{equation}}
\def\bea{\begin{eqnarray}}
\def\eea{\end{eqnarray}}

\begin{document}
\title{The Aharonov-Bohm Effect in Noncommutative  Quantum Mechanics}

\author{Kang Li$^{a,c}$}
\email{kangli@hztc.edu.cn}

\author{Sayipjamal Dulat$^{b,c}$}
\email{sdulat@xju.edu.cn}

\affiliation{$^{a}$Department of Physics, Hangzhou Teachers
College, Hangzhou, 310036, China}

\affiliation{$^{b}$Department of Physics, Xinjiang University,
Urumqi, 830046, China}

\affiliation{${}^c$The Abdus Salam International Center for
Theoretical Physics, Trieste, Italy}

\begin{abstract}
The Aharonov-Bohm (AB) effect in non-commutative quantum mechanics
(NCQM) is studied.  First, by introducing a shift for the magnetic
vector potential we give the Schr$\ddot{o}$dinger equations in the
presence of a magnetic field on NC space and NC phase space,
respectively. Then by solving the Schr$\ddot{o}$dinger equations,
we obtain the Aharonov-Bohm (AB) phase on NC space  and NC phase
space, respectively.

PACS number(s): 11.10.Nx, 03.65.-w
\end{abstract}

\maketitle
\section{Introduction: NC space and NC phase space}
Recently, there has been much interest in the study of physics on
noncommutative(NC) space\cite{SW}-\cite{Scho}, not only because
the NC space is necessary when one studies the low energy
effective theory of D-brane with B field background, but also
because in the very tiny string scale or at very high energy
situation, the effects of non commutativity of both space-space
and momentum-momentum may appear. There are many papers devoted to
the study of various aspects of quantum mechanics
\cite{DJT}-\cite{Likang}on noncommutative space with usual
(commutative) time coordinate.

In the usual $n$ dimensional commutative space , the coordinates and
momenta in quantum mechanics have the following commutation
relations:
\begin{eqnarray}
\label{Eq:cmr1}
 \begin{array}{l}
~[x_{i},x_{j}]=0,\\~ [p_{i},p_{j}]=0,\hspace{2cm} i,j = 1,2,
...,n,
\\~ [x_{i},p_{j}]=i \hbar\delta_{ij}.
\end{array}
\end{eqnarray}
At very tiny scales, say string scale,  not only space-momentum
does not commute, but also space-space  may not commute anymore.
Therefore the NC space is a space where coordinate and momentum
operators satisfy the following commutation relations
\begin{equation}\label{Eq:nmr2}
~[\hat{x}_{i},\hat{x}_{j}]=i\Theta_{ij},~~~
[\hat{p}_{i},\hat{p}_{j}]=0,~~~[\hat{x}_{i},\hat{p}_{j}]=i
\hbar\delta_{ij},
\end{equation}
where $\hat{x}_i$ and $\hat{p}_i$ are the coordinate  and momentum
operators on NC space. Ref.\cite{Likang} showed that
$\hat{p}_i=p_i$ and  $\hat{x}_i$  have the representation form
\begin{equation}\label{Eq:Rep.1}
 \hat{x}_{i}=  x_{i}-\frac{1}{2\hbar }\Theta_{ij}p_{j}, \hspace{1cm} i,j =
 1,2,...,n.
\end{equation}

The case of both space-space and momentum-momentum noncommuting
\cite{zhang}\cite{Likang} is different from the case of only
space-space noncommuting.  Thus the NC phase space is a space
where momentum operator in Eq. (\ref{Eq:nmr2}) satisfies the
following commutation relations
\begin{equation}\label{Eq:nmr3}
[\hat{p}_{i},\hat{p}_{j}]=i\bar{\Theta}_{ij},\hspace{2cm} i,j =
1,2,...,n.
\end{equation}
Here $\{\Theta_{ij}\}$ and $\{\bar{\Theta}_{ij}\}$  are totally
antisymmetric matrices which represent the noncommutative property
of the coordinate and momentum on noncommutative space and phase
space, respectively, and play analogous role to $\hbar$ in the usual
quantum mechanics. On NC phase space the representations of
$\hat{x}$ and $\hat{p}$ in term of $x$ and $p$ were given in
ref.\cite{Likang} as follows
\begin{equation}
\label{Eq:Rep.4}
 \begin{array}{ll}
 \hat{x}_{i}&= \alpha x_{i}-\frac{1}{2\hbar\alpha}\Theta_{ij}p_{j},\\
 ~&~\\
 \hat{p}_{i}&=\alpha p_{i}+\frac{1}{2\hbar\alpha}\bar{\Theta}_{ij}x_{j}, \hspace{1cm} i,j =
 1,2,...,n.
\end{array}
\end{equation}
The  $\alpha$  here is a scaling constant related to the
noncommutativity of phase space.When $\bar{\Theta}=0$, it leads
$\alpha =1$\cite{Likang}, the NC phase space returns to the NC
space, which is extensively studied in the literature, where the
space-space is non-commuting, while momentum-momentum is
commuting.

Given the NC space or NC phase space, one should study its physical
consequences. It appears that the most natural places to search the
noncommutativity effects are simple quantum mechanics (QM) system.
So far many interesting topics in NCQM such as hydrogen atom
spectrum in an external magnetic field \cite{nair,CST2} and
Aharonov-Bohm(AB) effect \cite{CST1} in the presence  of the
magnetic field, as well as the Aharonov-Casher effects \cite{MZ}
have been studied extensively. The  purpose of this paper is to do
further study on the Aharonov-Bohm effect on NC space and NC phase
space, respectively, where the space-space noncommutativity and
 both space-space and momentum-momentum noncommutativity could
produce additional phase difference.

This paper is organized as follows:  In section 2, we study the
Aharonov-Bohm  effect on NC space. First, the Schr$\ddot{o}$dinger
equation  in the presence of magnetic field is given, and the
magnetic Aharonov-Bohm phase expression is derived. In two
dimensions, our result agrees with the result of Ref. \cite{CST1}.
The general AB phase on NC space is also given in the presence of
 electromagnetic field.
 In section 3, we
investigate the Aharonov-Bohm effect on NC phase space. By solving
the Schr$\ddot{o}$dinger equation  in the presence of magnetic
field, the additional AB phase related to the momentum-momentum
noncommutativity is obtained explicitly.  Conclusions and some
remarks are given in  section 4.

\section{ The Aharonov-Bohm  effect on NC space}

Let $H(x,p)$ be the Hamiltonian operator of the usual quantum
system, then the static Schr$\ddot{o}$dinger equation on NC space
is usually written as
\begin{equation}\label{sdeq1}
 H(x,p)\ast\psi = E\psi,
\end{equation}
where the  Moyal-Weyl (or star) product between two functions is
defined by,
\begin{equation}\label{star}
(f  \ast g)(x) = e^{  \frac{i}{2}
 \Theta_{ij} \partial_{x_i} \partial_{x_j}
 }f(x_i)g(x_j)  = f(x)g(x)
 + \frac{i}{2}\Theta_{ij} \partial_i f \partial_j
 g\big|_{x_i=x_j},
\end{equation}
here $f(x)$ and $g(x)$ are two arbitrary functions. On NC space
the star product can be replaced by a Bopp's shift \cite{CFZ},
i.e. the star product can be changed into the ordinary product by
replacing $H(x,p)$ with  $H(\hat{x},\hat{p})$. Thus the
Schr$\ddot{o}$dinger equation can be written as,
\begin{equation}\label{GBShift}
H(\hat{x}_i,\hat{p}_i)\psi=H(
x_{i}-\frac{1}{2\hbar}\Theta_{ij}p_{j},p_{i})\psi = E\psi.
\end{equation}
Here $x_i$ and $p_i$ are coordinate and momentum operators in
usual quantum mechanics. Thus the Eq.(\ref{GBShift}) is actually
defined on commutative space, and the noncommutative effects can
be evaluated through the $\Theta$ related terms.  Note that the
$\Theta$ term always can be treated as a perturbation in QM ,
since $\Theta_{ij}<<1$.

When magnetic field is involved, the Schr$\ddot{o}$dinger equation
(\ref{sdeq1}) becomes
\begin{equation}\label{sdeq2}
H( x_i,p_i,A_i)\ast\psi = E\psi.
\end{equation}
To replace the star product in Eq.(\ref{sdeq2}) with a usual
product, first we need to replace $x_i$ and $ p_i$
 with  a Bopp's shift, then we also need to replace the
 vector potential $A_i$  with a  shift  given as follows
\begin{eqnarray}\label{L-D}
 A_i &\rightarrow& A_i +\frac{1}{2}\Theta_{lj}p_l\partial_j
 A_i,
\end{eqnarray}
 Thus the Schr$\ddot{o}$dinger
Eq.(\ref{sdeq2}) in the presence of magnetic field becomes
\begin{equation}\label{sdeq3}
 H\big( x_{i}-\frac{1}{2\hbar}\Theta_{ij}p_{j},p_{i},
  A_i +\frac{1}{2}\Theta_{lj}p_l\partial_j A_i \big)\psi =
E\psi.
\end{equation}
We should emphasize that the Bopp's shift and the shift Eq.
(\ref{L-D}) are equivalent to the star product in the
Schr$\ddot{o}$dinger Eq.(\ref{sdeq2}).

Now let us consider a particle of mass $m$ and charge $q$ moving
in a magnetic field with magnetic potential $A_i$, then the
Schr$\ddot{o}$dinger equation is (we choose unit of  $\hbar=c=1$),

\begin{equation}\label{sdeq3}
\frac{1}{2m} \Big(p_i- q A_i  - \frac{1}{2}q\Theta_{lj}p_l\partial_j
A_i \Big)^2\psi = E\psi.
\end{equation}
In an analogous way as in usual quantum mechanics, the solution
for (\ref{sdeq3}) reads
\begin{equation}\label{solution1}
\psi=\psi_0 \exp \big[iq\int_{x_0}^x(  A_i
+\frac{1}{2}\Theta_{lj}p_l\partial_j A_i)dx_i\big],
\end{equation}
where $\psi_0$ is the solution of (\ref{sdeq3}) when $A_i=0$. The
phase term of (\ref{solution1}) is so called AB phase. If we
consider a charged particle to pass a double slits,  then  the
integral runs from the source $x_0$ through one of the two slits to
the screen $x$, the coherent pattern will depend on the phase
difference of two paths. Thus the total phase shift for the AB
effect is
\begin{equation}\label{AB-sphase1}
\Delta \Phi_{AB}=\delta \Phi_0 + \delta\Phi_\theta^{NC} = iq\oint
 A_idx_i +\frac{iq}{2} \oint\Theta_{lj}(m v_l + q A_l)\partial_j A_i dx_i,
\end{equation}
 where the relation  $m v_l=p_l -q A_l+O(\Theta)$ has been
applied \footnote{By Eq.(\ref{sdeq3}), one writes the velocity
operator on NC space  as $v_l=\frac{\partial H}{\partial
p_l}=\frac{1}{m}(p_l-qA_l
-\frac{1}{2}q\Theta_{ij}p_i\partial_jA_l-\frac{1}{2}q\Theta_{lj}(p_i-qA_i)\partial_jA_i+O(\Theta^2))=\frac{1}{m}(p_l-qA_l
+O(\Theta)).$}, and we omitted the second order terms of the
$\Theta$; the first term is the AB phase in usual quantum
mechanics, the second term is the correction to the usual AB phase
due to space-space noncommutativity; the line integral runs from
the source through one of the two slits to the screen and returns
to the source through the other slit.

 In three dimensional NC space, i.e. $i,j=1,2,3$, we can
define a vector $\mathbf{\theta}=(\theta_1,\theta_2,\theta_3)$
with $\theta_i$ satisfies $\Theta_{ij}=\epsilon_{ijk}\theta_k$, or
 $\theta_i=\frac{1}{2}\epsilon_{ijk}\Theta_{jk}$. Then the second and third terms in Eq.(\ref{AB-sphase1})
 have the form

\begin{equation}\label{1}
 \frac{i }{2}q\oint \Theta_{lj} m v_l \partial_j A_i dx_i=\frac{i}{2}q
\oint \epsilon_{lmk}\theta_k m v_l\partial_j A_i dx_i =
\frac{i}{2}q m \oint\mathbf{\theta}\cdot (\mathbf{v}\times \nabla
A_i)dx_i,
\end{equation}
and
\begin{equation}\label{2}
\frac{i}{2}q^2\oint\Theta_{lj}A_l\partial_j A_i
dx_i=\frac{i}{2}q^2 \oint \epsilon_{lmk}\theta_k A_l\partial_j A_i
dx_i=\frac{i}{2}q^2 \oint \mathbf{\theta}\cdot(\mathbf{A
}\times\nabla A_i ) dx_i.
\end{equation}
Using Eqs. (\ref{1}) and (\ref{2}), we can write the AB phase as

\begin{equation}\label{AB-sphase2}
\Delta \Phi_{AB}=iq\oint  A_idx_i +\frac{i}{2}q \oint\big( m
 \mathbf{\theta} \cdot(\mathbf{v}\times \nabla A_i)+
q \mathbf{\theta}\cdot(\mathbf{A }\times\nabla A_i )\big)dx_i.
\end{equation}

In two dimensional NC plane $(i,j=1,2)$, if we consider an
electron $(q=-e)$ moving in a magnetic field,  then the vector
$\mathbf{\theta}$
 defined above just has the third component $\theta_3$  and
$\Theta_{ij}=\theta_3 \epsilon_{ij}$,
$\epsilon_{12}=-\epsilon_{21}=1, \epsilon_{11}=\epsilon_{22}=0$,
then we have
\begin{equation}\label{AB-sphase3}
\Delta \Phi_{AB}=-ie\oint  A_idx_i - \frac{i}{2}e
\mathbf{\theta}_3\oint\big( m
  (\mathbf{v}\times \nabla A_i)_3 - e(\mathbf{A }\times\nabla A_i )_3 \big)dx_i.
\end{equation}
We should emphasize that, in two dimensional NC plane, our result
(\ref{AB-sphase3}) is exactly the same as  in Ref. \cite{CST1}.

 The AB phase expression (\ref{AB-sphase1}) give us a hint that when a charged particle moves in
 an electromagnetic field with four dimensional potential $A_\mu$,
   then the corresponding AB phase will have the following
  general expression,
\begin{equation}\label{AB-sphase4}
\Delta \Phi_{AB}=iq\oint ( A_\mu
+\frac{1}{2}\Theta_{\alpha\beta}(mv_\alpha +q A_\alpha)
\partial_\beta A_\mu )dx^\mu .
\end{equation}
The second term is the consequence of space-space
non-commutativity.

\section{The Aharonov-Bohm  effect on NC Phase space }

The Bose-Einstein statistics in NCQM requires both space-space and
momentum-momentum non-commutativity. Thus we should also consider
the momentum-momentum non-commutativity when we deal with
 physical problems. On NC phase space the non-commuting
coordinates $\hat{x_i}$ and momentum $\hat{p_i}$ were given in Eq.
(\ref{Eq:Rep.4}). On NC phase space the star product in Eqs.
(\ref{star}) becomes,
\begin{equation}
(f  \ast g)(x,p) = e^{ \frac{i}{2\alpha^2}
 \Theta_{ij} \partial_i^x \partial_j^x+\frac{i}{2\alpha^2}\bar{\Theta}_{ij} \partial_i^p
 \partial_j^p}
 f(x,p)g(x,p)  = f(x,p)g(x,p)
 + \frac{i}{2\alpha^2}\Theta_{ij} \partial_i^x f \partial_j^x g\big|_{x_i=x_j}
 + \frac{i}{2\alpha^2}\bar{\Theta}_{ij} \partial_i^p f \partial_j^p g\big|_{p_i=p_j}.
\end{equation}
To replace the star product in Schr$\ddot{o}$dinger
Eq.(\ref{sdeq2}) with a usual product, first we need to replace
$x_i$ and $ p_i$   with a generalized Bopp's shift as
\begin{eqnarray}\label{gbshift1}
x_i\rightarrow x_{i}-\frac{1}{2\hbar\alpha^2}\Theta_{ij}p_{j},\nonumber\\
p_i\rightarrow p_{i}+\frac{1}{2\hbar\alpha^2}\bar{\Theta}_{ij}x_{j},
\end{eqnarray}
and then we need to replace $A_i$ with the generalized shift as
\begin{eqnarray}\label{L-D1}
A_i \rightarrow \alpha A_i
+\frac{1}{2\alpha}\Theta_{lj}p_l\partial_j
 A_i.
\end{eqnarray}
 Thus on NC phase space the Schr$\ddot{o}$dinger Eq.(\ref{sdeq2})
becomes,
\begin{equation}\label{sdeq4}
H\Big(  x_{i}-\frac{1}{2\hbar\alpha^2}\Theta_{ij}p_{j},
p_{i}+\frac{1}{2\hbar\alpha^2}\bar{\Theta}_{ij}x_{j}, A_i
+\frac{1}{2\alpha^2}\Theta_{lj}p_l\partial_j A_i \Big)\psi = E\psi.
\end{equation}
 One may note that the Eq.(\ref{gbshift1}) is different from
the Eq.(\ref{Eq:Rep.4}), by  Eq.(\ref{gbshift1}), the other
physical quantities may also be shifted, for example, mass may be
replaced with $m\rightarrow m/\alpha^2 $ and the electric charge
$q$ may be replaced with $q/\alpha$.

Here, again, we consider a particle of mass $m$ and electric charge
$q$ moving in a magnetic field.  On NC phase space, the Hamiltonian
have the form,
\begin{eqnarray}
\hat{H}=\frac{1}{2m} \Big(\alpha
p_i+\frac{1}{2\alpha}\bar{\Theta}_{ij}x_j- q (\alpha A_i
+\frac{1}{2\alpha}\Theta_{lj}p_l\partial_j A_i)
\Big)^2\nonumber\\
=\frac{1}{2m'} \Big( p_i+\frac{1}{2\alpha^2 }\bar{\Theta}_{ij}x_j-
q(A_i + \frac{1}{2\alpha^2}\Theta_{lj}p_l\partial_j A_i) \Big)^2,
\end{eqnarray}
with  $m'=m/\alpha^2  $. Thus the total phase shift for the AB
effect including the contribution due to both space-space and
momentum-momentum non-commutativity  on 3-dimensional NC phase
space is \footnote{ In a similar way as in  NC space, we have the
relation $m'v_l==p_l -q A_l+O(\Theta)+O(\bar{\Theta})$ on NC phase
space, and  we omitted the second order terms of $\Theta$ and
$\bar{\Theta}$ in Eq. ({\ref{AB-phase2}}).}
\begin{eqnarray}\label{AB-phase2}
\Delta \Phi_{AB}= \delta\Phi_{NCPS}&=&iq\oint A_idx_i
+\frac{iq}{2\alpha^2} \oint\big( m'
 \mathbf{\theta} \cdot(\mathbf{v}\times \nabla A_i)+
 q\mathbf{\theta}\cdot(\mathbf{A }\times\nabla A_i )\big)dx_i  - \frac{i}{2\alpha^2
 }\oint\bar{\Theta}_{ij}x_jdx_i\nonumber\\
&=&iq\oint A_idx_i +\frac{i}{2}q \oint\big( m
 \mathbf{\theta} \cdot(\mathbf{v}\times \nabla A_i)+ q\mathbf{\theta}\cdot(\mathbf{A }\times\nabla A_i\big)dx_i
+\delta\Phi^{NCPS}_{\bar\theta}.
\end{eqnarray}
Where the $\delta\Phi^{NCPS}_{\bar\theta}$ is the first order
modification term due to momentum-momentum non-commutativity, and
it has the form
\begin{eqnarray}\label{AB-phase3}
\delta\Phi^{NCPS}_{\bar\theta}= -\frac{i}{2\alpha^2 }\oint
\bar{\Theta}_{ij}x_jdx_i   +  \frac{i}{2\alpha^2}(1-\alpha^2 )
\oint q\mathbf{\theta}\cdot(\mathbf{A }\times\nabla A_i\big)dx_i
 + \frac{i}{2\alpha^4}(1-\alpha^4 )q \oint\big( m
 \mathbf{\theta} \cdot(\mathbf{v}\times \nabla A_i)dx_i.
\end{eqnarray}
It is obvious  from (\ref{AB-phase3}) that, when $\alpha=1$, then
we have $\bar{\Theta}_{ij}=0$ as well as
$\delta\Phi^{NCPS}_{\bar\theta} =0$, so the AB phase returns to
its expression Eq.(\ref{AB-sphase2}) on  NC space.

\section{Conclusion Remarks}

In this article we study the Aharonov-Bohm effect in NCQM. The
consideration of the NC space(NC phase space) produces an
additional phase difference. In order to obtain the NC space
correction to the usual Aharonov-Bohm phase difference, in section
two, first, we give the Schr$\ddot{o}$dinger equation  in the
presence of magnetic field, by solving the equation we derive the
magnetic Aharonov-Bohm phase expression. Note that the
non-commutative effects of the space (phase space) in the usual
Schr$\ddot{o}$dinger equation can be realized in two steps. First
step is to replace the coordinate and momentum operators with a so
called Bopp's (generalized Bopp's) shift, and then to replace the
magnetic potential $\mathbf{A}$ with a special shift which we
defined in Eq.(\ref{L-D}) in our paper. It is worth to mention
that, on NC plane, our result (\ref{AB-sphase3}) coincides with
the result of Ref. \cite{CST1}.  In order to obtain the NC phase
space correction to the usual Aharonov-Bohm phase difference, in
section 3, we solve the Schr$\ddot{o}$dinger equation in the
presence of magnetic field and obtain the magnetic Aharonov-Bohm
phase expression. Especially the new term
$\delta\Phi^{NCPS}_{\bar\theta}$ which comes from the
momentum-momentum noncommutativity is given explicitly.

The method we employed in this paper may also be applied to the
other related physical problems on NC space and NC phase space. For
example, the Aharonov-Casher effect in NC quantum mechanics. Further
study on the related topics will be reported in our forthcoming
paper. \vskip 1cm

{\bf{Acknowledgments:}} This paper is completed during our visit
to the high energy section of the Abdus Salam International Centre
for Theoretical Physics (ICTP). We would like to thank Prof. S.
Randjbar-Daemi for his kind invitation and warm hospitality during
our visit at the ICTP. This work is supported in part by the
National Natural Science Foundation of China (90303003, 10575026,
and 10465004). The authors also grateful to the support from the
Abdus Salam ICTP, Trieste, Italy.

\end{document}